\documentclass[12pt,preprint]{aastex}

\newcommand{\der}[2][\;\;]{\ensuremath{ \frac{d{#1}}{d{#2}} }}
\newcommand{\dern}[3][\;\;]{\ensuremath{ \frac{d^{#3}{#1}}{d{#2}^{#3}} }}
\newcommand{\dpar}[2][\;\;]{\ensuremath{ \frac{\partial{#1}}{\partial{#2}} }}
\newcommand{\dparn}[3][\;\;]{\ensuremath{ \frac{\partial^{#3}{#1}}{\partial{#2}^{#3}} }}

\newcommand{\e}{{e}}

\newcommand{\bvec}[1]{{\mbox{{\boldmath$#1$}}}} 
\newcommand{\unitv}[1]{\bvec{\hat{#1}}}

\newcommand{\eqnref}[1]{(\ref{#1})}

\begin{document}

\title{Kink Oscillations of a Curved, Gravitationally Stratified Coronal Loop }

\author{Bradley W. Hindman}
\affil{JILA and Department of Astrophysical and Planetary Sciences,
University of Colorado, Boulder, CO~80309-0440, USA}

\author{Rekha Jain}
\affil{Applied Mathematics Department, University of Sheffield, Sheffield S3 7RH, UK}

\email{hindman@solarz.colorado.edu}


\begin{abstract}

Loops of magnetic field in the corona are observed to oscillate and these oscillations
have been posited to be the superposition of resonant kink waves, trapped between
the loop's footpoints in the photosphere. To date, most analyses of these oscillations
have concentrated on calculating the frequency shifts that result from spatial
variation in the kink wave speed (produced primarily by stratification in the density
and magnetic field strength). Further, most have ignored gravity and treated the
loop as a straight tube. Here we ignore spatial variation in the wave speed, but
self-consistently include the effects of gravity and loop curvature in both the
equilibrium loop model and in the wave equation that governs the propagation of
the kink waves that live upon the loop. We model a coronal loop as an isolated,
thin, magnetic fibril that is anchored at two points in the photosphere. The equilibrium
shape of the loop is determined by a balance between magnetic buoyancy and magnetic
tension, which is characterized by a Magnetic Bond Number $\epsilon$ that is typically
small $\left|\epsilon\right| << 1$. This balance produces a loop that has a variable
radius of curvature, with the legs being relatively straight and the apex of the
loop the most curved. Further, a loop becomes unstable to buoyant rise if the footpoint
separation becomes larger than $2 \pi$ times the coronal pressure scale height. The
resonant kink waves of such a loop come in two polarizations that are decoupled from
each other: waves with motion completely within the plane of the loop (normal oscillations
or ``vertical modes") and waves with motions that are completely horizontal, perpendicular
to the plane of the loop (binormal oscillations or ``horizonal modes"). We solve for
the eigensolutions of both polarizations using perturbation theory for small Magnetic
Bond Number. For modes of the same order, normal oscillations have smaller eigenfrequencies
than binormal oscillations. The additional forces of buoyancy and magnetic tension
from the curvature of the loop increase and decrease the mode frequencies, respectively.
The ratio of the frequencies of the first overtone to the fundamental mode---a common
diagnostic used to assess the stratification of the wave speed along the loop---is
modified by the inclusion of buoyancy and curvature. We find that the normal polarization
possesses a frequency ratio that exceeds the canonical value of 2, whereas the binormal
polarization has a ratio less than 2. 

\end{abstract}

\keywords{MHD --- waves --- Sun: Corona --- Sun: magnetic fields --- Sun: oscillations}


\section{Introduction}
\label{sec:introduction}
\setcounter{equation}{0}

Initially through TRACE and EIT observations, and now ongoing with AIA imagery
\citep[e.g.,][]{White:2012}, the loops of magnetic field that often overlie magnetic
active regions are seen to oscillate with frequencies of typically 2--4 mHz
\citep[e.g.,][]{Schrijver:1999, Verwichte:2004, vanDoorsselaere:2007}. The attribution
of these oscillations to the modes of a one-dimensional cavity, and the observational
identification of overtone frequencies of this cavity, has given birth to the field
of coronal seismology (see Nakariakov \& Verwichte 2005 and the references therein).
The promise offered by observable resonant oscillation frequencies has raised hopes
for determining physical parameters of coronal loops which cannnot otherwise be measured
directly. 

Although multiple claims have been made for the observation of sausage waves
in post-flare coronal loops \citep[e.g.,][]{Srivastava:2008, Nakariakov:2005},
the majority of the observed oscillations useful for seismological purposes
are kink oscillations \citep{Schrijver:1999, Schrijver:2002, Aschwanden:1999,
Aschwanden:2002, Verwichte:2004}. The dispersion in the observed overtones,
compared to theoretical predictions, have been used mainly to estimate the
mean magnetic field strength \citep[e.g.,][]{Nakariakov:2001}, it's longitudinal
gradient \citep{Verth:2008, Ruderman:2008}, or the density stratification
inside and outside the loop \citep{Donnelly:2006, Diaz:2004, Diaz:2007, McEwan:2008,
Orza:2012, vanDoorsselaere:2007, Goossens:2006}. Observational efforts are
now focused on improving the measurement of oscillation eigenfrequencies by
reducing the observational errors and by analyzing additional loops. A
relevant overview of observational measurements can be found in \cite{Andries:2009}.

On the theoretical side, advancements are steadily being made by including additional
physical effects with the aim of developing a general model of coronal loops and their
oscillations. Most of this work is based on the pioneering calculations of \cite{Edwin:1983},
which were presciently performed prior to the first observations of coronal loop waves.
This work revealed the MHD wave modes of a translationally invariant magnetic cylinder.
Since then \cite{Diaz:2004} and \cite{Andries:2005} have investigated modifications
to the eigenfrequencies that arise from density stratification along the loop.
The expected dispersion has been compared to measured values with the goal
of constraining coronal loop models \citep[see also][]{Goossens:2006, Andries:2005, McEwan:2008}.
Further, the quantitative effects of loop geometry were investigated by
\cite{Dymova:2006} and \cite{Verth:2008}, with particular focus on the curvature
of the loop axis. For example, \cite{vanDoorsselaere:2004} have shown
that the frequency shift induced by curvature appears at second order in
the ratio of tube radius to the loop length, which is  believed to be quite small
for the majority of loops. They and others have also suggested that
since the eigenfrequencies for both horizontal and vertical \citep{Wang:2004}
polarizations are identical, there is no distinction between the oscillations
with different polarization \citep[e.g.,][]{Terradas:2006}, unless of course
the tube is not locally axisymmetric \citep{Ruderman:2003}. 

Here, our goal is to self-consistently include the effects of gravity and curvature,
both in the establishment of the static equilibrium of a coronal loop and in the
propagation of kink waves along the loop. We accomplish both of these goals by
exploiting the thin-tube approximation, whereby lateral variation across the tube
is ignored. Under this assumption the equilibrium shape of the coronal loop is
established by a balance between the forces of magnetic buoyancy and magnetic
tension. The eigenmode problem is reduced to a one-dimensional wave equation
which we solve through perturbation theory. The paper is organized as follows:
in Section 2 we present the equation of motion appropriate for a thin magnetic fibril.
In Section 3 we derive the equilibrium height and shape of the loop, while Section 4
details the solution to the kink wave eigenmode problem. In Section 5, we discuss
the significance of our findings and the implications of our base assumptions. Finally,
we state our findings and conclusions in Section 6.


\section{Equation of Motion}
\label{sec:EqMotion}

We will treat a coronal loop as a thin magnetic fibril embedded within a field-free
coronal atmosphere. We recognize that both of these assumptions are suspect. Even
though the thin tube approximation may only apply to a subset of coronal loops,
it provides mathematical tractability. The assumption that the loop is an isolated
magnetic structure is probably never strictly valid, as the entirety of the solar
corona is magnetically dominated and most loops are an integral part of larger
magnetic structures. In fact, most loops are probably just selectively illuminated
field lines amongst many that form magnetic arcades. Despite these objections, we
adopt an isolated fibril model for convenience. The consequences of these assumptions
are discussed in more detail in Section~\ref{sec:Discussion}.

The forces acting on an isolated, thin magnetic fibril have been previously derived
by \cite{Spruit:1981} through averaging the MHD momentum equation over the cross-sectional
area of the tube. In terms of the acceleration of the tube, the resulting equation of
motion is given by

\begin{equation}
\label{eqn:Spruit}
	\der[\bvec{v}]{t} = \left[\unitv{s} \cdot \left( -\frac{1}{\rho} \bvec{\nabla}p + \bvec{g}\right) \right] \unitv{s}
		+ c^2 \bvec{k} + D \left(\unitv{s} \times \bvec{g}\right) \times \unitv{s} \; ,
\end{equation}

\noindent where all fluid properties represent cross-sectional averages and $\unitv{s}$
and $\bvec{k}$ are the instantaneous tangent unit vector and curvature vector of the
tube's axis. The gravitational acceleration is $\bvec{g}$ and $p$ is the gas pressure.
The longitudinal forces (those in the $\unitv{s}$ direction) are just the pressure and
gravitational forces that act parallel to the magnetic field. The last two terms on the
right-hand side are the transverse forces of magnetic tension and buoyancy with $c$ being
the kink wave speed and $D$ the tube's fractional overdensity compared to its surroundings,

\begin{eqnarray}
	c^2 &\equiv& \frac{B^2}{4\pi \left( \rho + \rho_\e \right)} \; ,
\\ \nonumber \\
	D &\equiv& \frac{\rho - \rho_\e}{\rho + \rho_\e} \; .
\label{eqn:D}
\end{eqnarray}

\noindent The densities $\rho$ and $\rho_\e$ are, respectively, the mass density
internal and external to the tube and $B$ is the axial magnetic field strength
of the fibril.

The transverse forces (tension and buoyancy) in equation~\eqnref{eqn:Spruit} include
the effects of ``enhanced inertia," which represents the backreaction caused by the
external medium when transverse motions of the tube move surrounding fluid out of
the way. The form of the enhanced inertia is the same as propounded by \cite{Spruit:1981},
whereby the density in the inertial term is simply augmented, $\rho \to \rho + \rho_\e$.
This form has inspired much controversy \citep[e.g.,][]{Choudhuri:1990, Cheng:1992, Fan:1994},
but faithfully reproduces the proper kink wave speed and is consistent with all alternate
formulations as long as (1) the background atmosphere lacks rotational shear and magnetic
field, (2) flow along the tube is not permitted, and (3) correction terms that are quadratic
in the fluid velocity are ignored.


\section{The Equilibrium Shape of the Loop}
\label{sec:Equilibrium}

\begin{figure*}%
        \epsscale{0.5}%
        \plotone{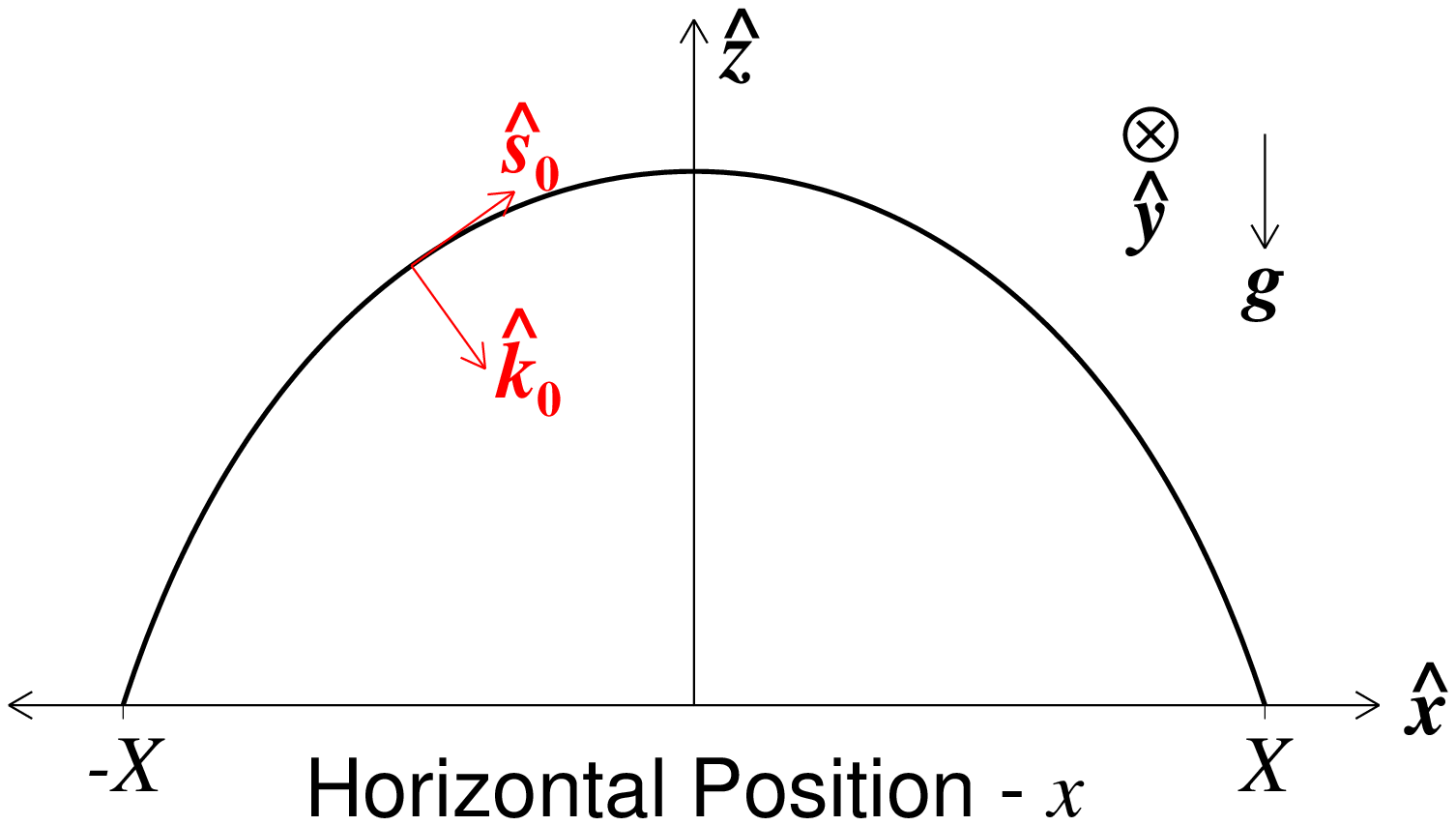}%
        \caption{\small The geometry of the coronal loop, with the loop shown in
black. The photosphere corresponds to the $x$--$y$ plane, while the loop is confined
to the $x$--$z$ plane. The local Frenet coordinates are indicated in red. The local
tangent vector is $\unitv{s}_0$ and the principle normal $\unitv{k}_0$ lies in the
direction of curvature. The binormal is everywhere constant and pointed in the
$\unitv{y}$ direction.
\label{fig:Geometry}}%

\end{figure*}

We neglect the spherical geometry of the solar atmosphere and assume that the corona
can be treated as plane-parallel with constant gravity $\bvec{g}$. We employ a
Cartesian coordinate system, with the $x$--$y$ plane corresponding to the photosphere
and the $z$ coordinate increasing upwards (i.e., $\bvec{g} = -g \unitv{z}$). The
equilibrium position of the coronal loop is confined to the $x$--$z$ plane with the
endpoints of the loop anchored in the photosphere at the coordinates ($\pm X$,0,0).
The height of the loop above the photosphere is given by the function $z_0(x)$, such
that the loop's axis is traced by the position vector $\bvec{r}_0(x) = x \unitv{x} + z_0(x) \unitv{z}$.
Figure~\ref{fig:Geometry} provides an illustration of the geometry of the loop and
the coordinate system.

In addition to this Cartesian coordinate system, we define a local set of Frenet
coordinates for the equilibrium position of the loop. Let $s$ denote the arclength
along the loop measured from the footpoint located at $x=-X$. The longitudinal unit
vector, or the tangent vector that lies parallel to the loop's axis pointing in the direction
of increasing $s$, is denoted $\unitv{s}_0$. Within the $x$--$z$ plane and everywhere
perpendicular to $\unitv{s}_0$ is the principle normal $\unitv{k}_0$. We shall soon
see that for the loop model we adopt, only convex loops are stable. Therefore, the
curvature vector $\bvec{k}_0$ always points inward and downward in the direction of
the principle normal $\unitv{k}_0$, with a magnitude equal to the reciprocal of the
local radius of curvature $R_0^{-1}$ of the loop's axis. The third orthogonal direction
is given by the binormal unit vector $\unitv{b}_0 = \unitv{s}_0 \times \unitv{k}_0$,
which is everywhere constant $\unitv{b}_0 = \unitv{y}$. In what follows we will need
the standard geometrical equations which describe the Frenet unit vectors,
$\unitv{s}_0$ and $\unitv{k}_0$, and the radius of curvature $R_0$ in terms of the
loop height,

\begin{eqnarray}
\label{eqn:s0hat}
	\unitv{s}_0 &\equiv& \frac{\partial \bvec{r}_0}{\partial s} =
		\frac{\unitv{x} + z_0^\prime(x)\unitv{z}}{s^\prime(x)} \; ,
\\ \nonumber \\
	\bvec{k}_0 &\equiv& \frac{\partial \unitv{s}_0}{\partial s} =
		\frac{1}{R_0} \frac{z_0^\prime(x) \unitv{x} - \unitv{z}}{s^\prime(x)} \; ,
\label{eqn:k0} \\ \nonumber \\
	R_0(x) &\equiv& \frac{1}{\left|\bvec{k}_0\right|} =
		- \frac{\left[s^\prime(x)\right]^3}{z_0^{\prime\prime}(x)} \; ,
\label{eqn:Rcurv} \\ \nonumber \\
	s^\prime(x) &=& \sqrt{1 + \left[ z_0^{\prime}(x) \right]^2} \; ,
\label{eqn:sprime}
\end{eqnarray}

\noindent where primes denote differentiation with respect to the photospheric coordinate $x$.

Since a loop in a state of static equilibrium is confined to the $x$--$z$ plane,
none of the forces on the fibril have a component in the binormal $y$-direction
and we only need to consider two components of the equation of motion~\eqnref{eqn:Spruit},

\begin{eqnarray}
\label{eqn:equil_s}
	-\frac{1}{\rho_0} \dpar[p_0]{s} - g \unitv{z} \cdot \unitv{s}_0 &=& 0 \; ,
\\
	\frac{c_0^2}{R_0} - gD_0 \unitv{z} \cdot \unitv{k}_0 &=& 0 \; .
\label{eqn:equil_k}
\end{eqnarray}

\noindent In these two equations the subscript 0 indicates a background, equilibrium
quantity within the loop. The triple cross-product appearing in the buoyancy term in
equation~\eqnref{eqn:Spruit} has been reduced to a simpler form by making use
of the identities
$\left(\unitv{s}_0 \times \unitv{z}\right) \times \unitv{s}_0=\unitv{z}-(\unitv{z}\cdot\unitv{s}_0)\unitv{s}_0=(\unitv{z}\cdot\unitv{k}_0)\unitv{k}_0$.

Equation~\eqnref{eqn:equil_s} expresses the balance of forces in the tangential
or axial direction $\unitv{s}_0$, whereas equation~\eqnref{eqn:equil_k} describes
the balance in the transverse direction of the principle normal $\unitv{k}_0$. The
axial equation is simply hydrostatic balance along magnetic field lines which we
will satisfy trivially by assuming that the loop and the surrounding corona are
both isothermal. Furthermore, since the tube is thin and thermal diffusion wipes
out lateral variations, the temperatures inside and outside the tube are identical,
$T_0 = T_\e$. All densities and pressures (both gas and magnetic) therefore vary
exponentially with height $z$ with a common scale height $H = R_{\rm gas} T_0/g$.
A consequence of this thermal uniformity is that the kink speed $c_0$, the
plasma-parameter $\beta=8\pi p_0/B_0^2$, and the overdensity $D_0$ are all constants.

The shape of the loop $z_0(x)$ is constrained by the balance of the transverse
forces of magnetic tension and magnetic buoyancy, which is quantified by
equation \eqnref{eqn:equil_k}. Since the temperatures inside and outside the loop
are identical, and the external fluid is field free, the tube must be partially
evacuated ($D_0<0$) in order to ensure pressure continuity across the lateral
boundary of the flux tube. Thus, magnetic buoyancy tries to lift the loop higher
into the atmosphere and magnetic tension arising from the convex curvature of the
loop tries to hold it down. Since the kink speed $c_0$ is constant for our
isothermal model, all variation along the loop in the magnitude of the magnetic
tension force comes from changes in the radius of curvature $R_0$. Similarly,
the buoyancy only varies because the transverse direction $\unitv{k}_0$ rotates
relative to the vertical $\unitv{z}$.

If we insert equations~\eqnref{eqn:k0} and \eqnref{eqn:Rcurv} into equation~\eqnref{eqn:equil_k},
we obtain a nonlinear ODE for the height of the loop $z_0(x)$,

\begin{equation} \label{eqn:EqODE}
	\frac{z_0^{\prime\prime}}{1+\left(z_0^\prime\right)^2} = \frac{\epsilon}{X} \; ,
\end{equation}

\noindent where we have defined a Magnetic Bond Number,

\begin{equation} \label{eqn:epsilon}
	\epsilon \equiv \frac{g D_0 X}{c_0^2} \; .
\end{equation}

\noindent The Magnetic Bond Number can be written in a more recognizable form if
we multiply the numerator and denominator by the enhanced inertia $\rho+\rho_e$,

\begin{equation} \label{eqn:epsilon_variant}
	\epsilon = \frac{g (\rho - \rho_\e) X}{B^2/4 \pi} \; .
\end{equation}

\noindent In this expression the surface tension that normally appears in the
denominator of the Bond Number has been replaced by magnetic tension. The Magnetic
Bond Number $\epsilon$ is a signed, dimensionless number that embodies the relative
importance of the buoyancy and magnetic forces. The Magnetic Bond Number appears
in calculations of the Rayleigh-Taylor instability when one of the fluid layers
is filled with a horizontal field, providing the critical wavenumber below which
instability ensues, $k_c = -\epsilon / X$ \citep{Chandrasekhar:1961}.

For our isothermal model, $\epsilon$ is a constant and furthermore it can be expressed
as a ratio of length scales, in particular the ratio of the footpoint separation to
the pressure scale height $\epsilon = -X/2H$. This property is easily derived by
realizing that the combination of the perfect gas law and pressure continuity across
the lateral surface of the tube allows one to express the magnetic pressure inside
the tube as a ratio of the density difference to the fluid temperature. When this expression
is inserted into equation~\eqnref{eqn:epsilon_variant}, the density difference cancels
in the numerator and denominator, leaving just the pressure scale height in the denominator
(which is proportional to the temperature).

Since $\epsilon$ is a constant, equation~\eqnref{eqn:EqODE} can be integrated directly
to obtain the derivative of the height function $z_0^\prime(x)$. Similarly, this
expression for the derivative can be subsequently integrated to obtain the height
function itself $z_0(x)$. We choose the constants of integration such that $z_0(\pm X) = 0$
and $z_0^\prime(0) = 0$. A summary of the resulting properties of the loop is provided
below:
 
\begin{eqnarray}
	z_0(x) &=& \frac{X}{\epsilon} \ln\left(\frac{\cos \epsilon}{\cos\left(\epsilon x/X\right)}\right) \; ,
\label{eqn:z0} \\ \nonumber \\
	z_0^\prime(x) &=& \tan\left(\epsilon x/X\right) \; ,
\label{eqn:dz0} \\ \nonumber \\
	z_0^{\prime\prime}(x) &=& \frac{\epsilon}{X} \sec^2\left(\epsilon x/X\right) \; ,
\label{eqn:d2z0} \\ \nonumber \\
	R_0(x) &=& - \frac{X}{\epsilon} \sec\left(\epsilon x/X\right) \; ,
\label{eqn:R0} \\ \nonumber \\
	s(x) &=& \frac{X}{2\epsilon} \ln\left( \frac{1+\sin\epsilon}{1-\sin\left(\epsilon x/X\right)}
			\frac{1+\sin\left(\epsilon x/X\right)}{1-\sin\epsilon} \right) \; .
\label{eqn:pathlength}
\end{eqnarray}

\begin{figure*}%
        \epsscale{1.0}%
        \plotone{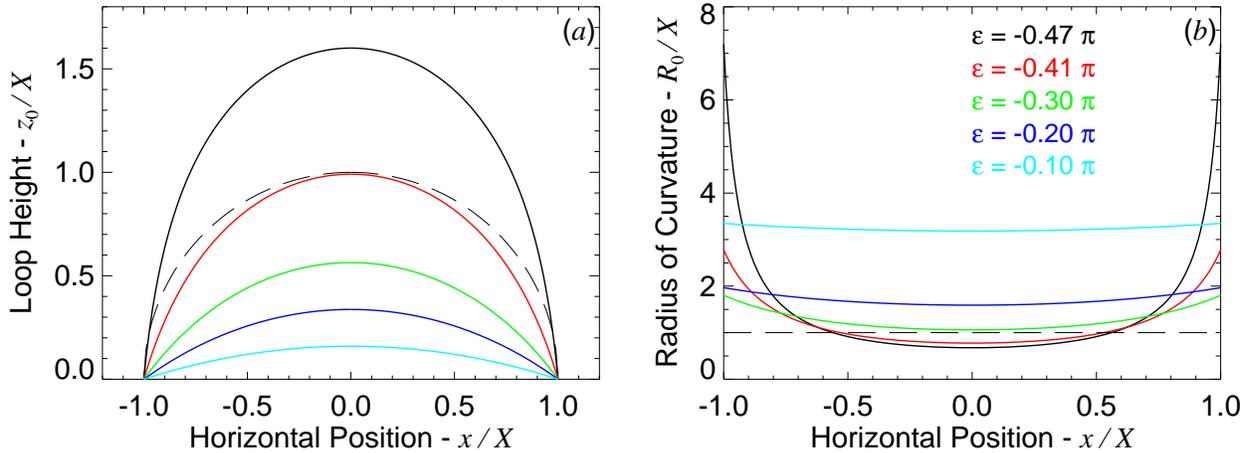}%
        \caption{\small Equilibrium ($a$) height and ($b$) radius of curvature
of coronal loop models with differing ratios of buoyancy to magnetic tension, i.e.,
differing values of the Magnetic Bond Number $\epsilon$. The value of $\epsilon$ associated
with each color is indicated in the right panel.  In both panels the dashed line
corresponds to a semi-circle with constant radius of curvature $R_0 = X$. Loops with
weak buoyancy ($\left|\epsilon\right|<< 1$) are flat with large radius of curvature
everywhere. Loops with substantial buoyancy ($\epsilon \approx - \pi/2$) are tall with
large radius of curvature in the legs and small radius of curvature at its apex.
Loops with positive $\epsilon$ or with $\epsilon < -\pi/2$ are unstable.
\label{fig:LoopModel}}%

\end{figure*}

\noindent The last of these equations was obtained by integrating equation~\eqnref{eqn:sprime},
with the constant of integration chosen such that $s(-X) = 0$. Figure~\ref{fig:LoopModel}
displays solutions for several different values of $\epsilon$. For reference, the
dashed curve corresponds to a semi-circle, clearly illustrating the variable radius
of curvature for the loops.

The length of the loop $L$ is obtained by inserting the footpoint position $x = X$ into
equation~\eqnref{eqn:pathlength},

\begin{equation} \label{eqn:Length}
	L = \frac{X}{\epsilon} \ln\left( \frac{1+\sin\epsilon}{1-\sin\epsilon} \right) \; .
\end{equation}

\noindent Two interesting limits of this equation exist. As $\epsilon \to 0$ the
length of the loop converges to the footpoint separation $L \to 2X$. This arises
because the loop becomes straight, flat, and confined to the photosphere.  As $\epsilon \to \pm \pi/2$
the loop length diverges logarithmically because the height of the loop grows without
bound. This can be seen by recognizing that the loop reaches its apex at its center
($x=0$), therefore achieving a maximum height

\begin{equation}
	z_{\rm apex} = z_0(0) = \frac{X}{\epsilon} \ln\left(\cos\epsilon\right) \; .
\end{equation}

\noindent One can easily see that $\epsilon = \pm\pi/2$ corresponds to a logarithmic
singularity in the height.

Clearly the height $z_0(x)$ should be a positive function for the range $x \in (-X,~X)$.
With a little thought, from equation~\eqnref{eqn:z0} we can see that this is only
possible if $\epsilon \in (-\pi/2,~0)$. Therefore, for constant $c_0$ only underdense
tubes ($D_0<0$) form stable loops and those loops are comprised of a single arch without
inflection points ($z_0^{\prime\prime} < 0$). Furthermore, for $z_0$ to remain positive,
$\epsilon$ must be bounded and an equilibrium is only possible if the magnetic tension
exceeds a threshhold,

\begin{equation}
	\frac{B_0^2}{4\pi} > \frac{B_c^2}{4\pi} = 2 \pi^{-1} g (\rho_\e - \rho) X \; .
\end{equation}

\noindent This condition for stability can also be used to constrain the footpoint
separation. Since, $\epsilon = -X/2H$, stability requires that the footpoint separation
be less than a critical value that depends on the corona's pressure scale height,

\begin{equation}
	2X < 2X_c = 2 \pi H \; .
\end{equation}

Typically, we do not need to be overly concerned about the limit of large buoyancy
forces. For most coronal loops the Magnetic Bond Number $\epsilon$ is quite small.
Using typical values of the footpoint separation $2X = 150$ Mm, kink wave speed $c_0 = 1$
Mm s$^{-1}$ \citep{Tomczyk:2007}, overdensity $D_0 = -1$ (appropriate for very small
plasma parameter $\beta$), and gravity $g = 2.7 \times 10^{-4}$ Mm s$^{-2}$, we find
from equation~\eqnref{eqn:epsilon}  that $\epsilon = -2\times 10^{-2}$. For small
values of $\epsilon$, the loop is short and low-lying with
$z_{\rm apex} = \left|\epsilon\right| X/2$. This geometry is realized because the
field lines exit the photosphere at an oblique angle---i.e., 
$z_0^\prime(\pm X) = \tan\epsilon$. As $\epsilon$ becomes small, the angle between
the fibril's axis and the photospheric surface approaches zero (see Figure~\ref{fig:LoopModel}).


\section{Kink Wave Oscillations}
\label{sec:KinkWaves}

Kink waves cause the axis of the fibril to undulate; therefore, we need to consider
wave-induced modulation of the local direction vectors, $\unitv{s}$ and $\unitv{k}$.
Let the instantaneous position vector of the loop be described as a time-varying
perturbation about a static equilibrium. We will consider two transverse displacements
of the axis, one displacement $\xi(s,t)$ polarized in the direction of the principle
normal $\unitv{k}_0$ (which is confined to the $x$--$z$ plane) and a perpendicular
displacement $\zeta(s,t)$ polarized in the binormal $\unitv{y}$ direction (i.e.,
horizontal and parallel to the photospheric plane),

\begin{equation}
	\bvec{r}(s,t) = \bvec{r}_0(s) + \xi(s,t) \ \unitv{k}_0(s) + \zeta(s,t) \ \unitv{y} \; .
\end{equation}

In the equation above, as before, the coordinate $s$ is the pathlength along the unperturbed
position of the tube. If we carefully consider the perturbations to the unit vectors,
we find to first order in the displacements

\begin{eqnarray}
	\unitv{s} &=& \unitv{s}_0 + \dpar[\xi]{s} \ \unitv{k}_0 + \dpar[\zeta]{s} \ \unitv{y} \; ,
\label{eqn:sdir} \\ \nonumber \\
	\bvec{k} &=& \frac{\unitv{k}_0}{R_0} - \frac{1}{R_0} \dpar[\xi]{s} \ \unitv{s}_0
		+ \left( \dparn[\xi]{s}{2} + \frac{\xi}{R_0^2} \right) \ \unitv{k}_0
		+ \dparn[\zeta]{s}{2} \unitv{y} \; ,
\label{eqn:kdir} \\ \nonumber \\
	R^{-1} &=& R_0^{-1} + \left( \dparn[\xi]{s}{2} + \frac{\xi}{R_0^2} \right)  \; .
\label{eqn:RcurvWave}
\end{eqnarray}

\noindent Since, the equilibrium configuration of the loop lacks curvature in the
$\unitv{y}$ direction, the radius of curvature does not depend on the $\zeta$ displacement,
at least to linear order. Because of this, the two polarizations decouple, and $\xi$
and $\zeta$ satisfy independent equations. These two wave equations are obtained by
noting that transverse oscillations of the thin fibril are largely incompressive;
therefore, the longitudinal component of the equation of motion, equation~\eqnref{eqn:Spruit},
can be ignored.

Differential equations describing the two polarizations of oscillation can be derived
by inserting equations~\eqnref{eqn:sdir}--\eqnref{eqn:RcurvWave} into equation~\eqnref{eqn:Spruit},
Fourier transforming in time, and linearizing in the displacements,

\begin{eqnarray}
\label{eqn:xieqn_raw}
	c_0^2 \left(\dern[]{s}{2} + \frac{1}{R_0^2}\right) \xi - g_\parallel D_0 \der[\xi]{s}
		+ \omega^2 \xi &=& 0 \; ,
\\ \nonumber \\
	c_0^2 \dern[\zeta]{s}{2} - g_\parallel D_0 \der[\zeta]{s}
		+ \omega^2 \zeta &=& 0 \; .
\label{eqn:zetaeqn_raw}
\end{eqnarray}

\noindent The influence of the geometry of the equilibrium loop model is felt only
through the radius of curvature $R_0$ and through the parallel component of gravity
$g_\parallel \equiv \bvec{g} \cdot \unitv{s}_0$. The terms proportional to $c_0^2$
represent magnetic tension, with the radius of curvature corresponding to the contribution
from the curvature of the equilibrium field lines. The terms with $g_\parallel$ are
buoyancy forces and the inertial terms are those that depend on the temporal frequency
$\omega$. These are quite general equations that describe oscillations for any loop
equilibrium---not just the isothermal model we consider here. In fact, except for the
inclusion of curvature, these are the same equations used by \cite{Jain:2012} to study
the effects of buoyancy and wave speed variations on mode frequencies. 

Let's first examine the tension terms in detail. When these two equations are compared,
we see that the curvature of the tube only appears in the equation for the normal
oscillations. This additional tension term from curvature typically has the opposite
sign from the tension arising from undulations. Thus, the two tension forces that
appear in the equation for normal oscillations act in opposition and we should expect
that curvature of the loop reduces the eigenfrequencies. Further, if we consider the
limit of small Magnetic Bond Number $\epsilon$, we can see from equation~\eqnref{eqn:R0}
that the radius of curvature is large $R_0 = O(\epsilon^{-1})$. Therefore, the curvature of
the tube should generate a second-order frequency shift $\Delta\omega/\omega = O(\epsilon^2)$.

If we insert $g_\parallel \equiv -g \unitv{z} \cdot \unitv{s}_0$ into equation~\eqnref{eqn:zetaeqn_raw},
and include the Magnetic Bond Number dependence directly, we get

\begin{equation}
	\dern[\zeta]{s}{2} + \frac{\epsilon}{X} \; \unitv{z}\cdot\unitv{s}_0 \; \der[\zeta]{s}
		+ \frac{\omega^2}{c_0^2} \zeta = 0 \; .
\end{equation}

\noindent A cursory examination of this equation would seem to indicate that for small
$\epsilon$, the frequency shift from buoyancy forces should be a first-order quantity.
However, this is incorrect. We have previously demonstrated that the equilibrium height
and shape of the loop sensitively depend on the Magnetic Bond Number, and in particular
as $\epsilon \to 0$ the loop becomes flat and horizontal. In fact, by using
equations~\eqnref{eqn:s0hat}, \eqnref{eqn:sprime}, and \eqnref{eqn:dz0} one finds

\begin{equation}
	\unitv{z}\cdot\unitv{s}_0 = \sin\left(\epsilon x / X\right) \; .
\end{equation}

\noindent Since $\sin\left(\epsilon x / X\right) = O(\epsilon)$ for small $\epsilon$, it
is now clear that buoyancy should cause frequency shifts that are second order in the
Magnetic Bond Number. This makes sense, as curvature also enters at second order and
buoyancy and curvature are balanced in the equilibrium configuration of the loop.

The quantities $g_\parallel$ and $R_0$ depend on the equilibrium shape of the loop, which
depends explicitly on the horizontal coordinate $x$---see
equations~\eqnref{eqn:z0}--\eqnref{eqn:pathlength}. Therefore, it is quite natural to make
a change of variable from the arclength variable $s$ to the horizontal $x$ coordinate. After
inserting the properties of the equilibrium, we find that the resulting ODEs lack first-derivative
terms,

\begin{eqnarray}
\label{eqn:xieqn}
	\dern[\xi]{x}{2} + \left[ \frac{\omega^2}{c_0^2} \ \frac{1}{\cos^2\left(\epsilon x/X\right)}
		+ \frac{\epsilon^2}{X^2} \right] \xi &=& 0 \; ,
\\ \nonumber \\
	\dern[\zeta]{x}{2} + \frac{\omega^2}{c_0^2} \ \frac{1}{\cos^2\left(\epsilon x/X\right)} \ \zeta &=& 0 \; .
\label{eqn:zetaeqn}
\end{eqnarray}

\noindent The absence of the first derivatives occurs because the tension and buoyancy forces
in the equilibrium are in exact balance.

Both of these equations have solutions in the form of Associated Legendre Functions,

\begin{eqnarray}
	\xi(x) = \left(\cos\theta\right)^{1/2} \left[ A_1 P_{1/2}^\mu\left(\sin\theta\right) 
		+ A_2 Q_{1/2}^\mu\left(\sin\theta\right) \right] \; ,
\\ \nonumber \\
	\zeta(x) = \left(\cos\theta\right)^{1/2} \left[ A_3 P_{-1/2}^\mu\left(\sin\theta\right) 
		+ A_4 Q_{-1/2}^\mu\left(\sin\theta\right) \right] \; ,
\end{eqnarray}

\noindent with

\begin{eqnarray}
	\theta(x) &\equiv& \epsilon \frac{x}{X} \; , 
\\ \nonumber \\
	\mu &\equiv& \left(\frac{1}{4} - \frac{\omega^2 c_0^2}{g^2 D_0^2}\right)^{1/2} \; .
\end{eqnarray}

\noindent The dependence on the frequency $\omega$ is buried in the upper index $\mu$
of the Associated Legendre Functions. Therefore, these solutions are not particularly
friendly, and applying boundary conditions to quantize the eigenfrequencies would require
a nontrivial numerical computation. We avoid this difficulty by realizing that the
Magnetic Bond Number $\epsilon$ is  a small dimensionless parameter, thus allowing
the use of perturbation theory.

The equations that describe oscillations in both the normal and binormal directions
have very similar form. In fact, a brief inspection of equations~\eqnref{eqn:xieqn}
and \eqnref{eqn:zetaeqn} reveals that the eigenfunctions of the two polarizations should
be identical up though second order in $\epsilon$. This means that the derivation of
the solutions for the two polarizations are similar; therefore, we will demonstrate
the perturbation analysis for only the normal component and simply state the result
for the binormal component. Assume that the eigenfunction and eigenfrequency can be
written as perturbation expansions,

\begin{eqnarray}
	\xi_n(x) &=& \Xi_n(x) + \epsilon^2 \delta\xi_n(x) + O(\epsilon^4) \; ,
\\ \nonumber \\
	\omega_n^2 &=& \Omega_n^2 + \epsilon^2 \delta\omega_n^2 + O(\epsilon^4) \; .
\end{eqnarray}

\noindent Insert these expansions into equation~\eqnref{eqn:xieqn} and solve in the usual
fashion for the zeroth-order eigenfunction and eigenfrequency, noting that the
cosine appearing in the second term can be expanded for small argument,

\begin{eqnarray}
	\Xi_n(x) &=& X \sin\left[\kappa_n (X-x) \right] \; ,
\\ \nonumber \\
	\Omega_n &=& \kappa_n c_0 = \frac{n\pi c_0}{2X} \; .
\label{eqn:Omega_n}
\end{eqnarray}

\noindent The boundary condition of vanishing displacement at the footpoints $\xi(\pm X) = 0$
has been applied.

If we now collect terms of the next order, we find the following inhomogeneous
equation for the perturbed eigenfunction,

\begin{equation}
	\left(\dern[]{x}{2} + \frac{\Omega_n^2}{c_0^2} \right) \delta\xi_n =
		-\left( \frac{\Omega_n^2}{c_0^2} \frac{x^2}{X^2}
			+ \frac{\delta\omega_n^2}{c_0^2} + \frac{1}{X^2} \right) \Xi_n \; .
\end{equation} 

\noindent The particular solution can be found by judicious guessing and by enforcing that
the perturbed eigenfunction has the same symmetry about the center of the loop as the
unperturbed eigenfunction,

\begin{equation}
	\delta\xi_n(x) = -\frac{x^2}{X^2} \Xi_n(x) + \left[ \frac{x^3}{6X^2}
		+ \frac{c_0^2}{\Omega_n^2} \frac{x}{2} \left(\frac{\delta\omega_n^2}{c_0^2}
			+ \frac{1}{2X^2}\right) \right] \Xi_n^\prime(x) \; .
\end{equation}

\noindent The perturbed eigefrequency is fixed by reinforcing that the eigenfunction
vanishes at the footpoints for all orders of expansion $\delta\xi_n(\pm X) = 0$,

\begin{equation}
	\delta\omega_n^2 = -\left( \frac{\Omega_n^2}{3} + \frac{c_0^2}{2X^2} \right) \; .
\end{equation}

If we now put all the pieces together, to second order in the small parameter $\epsilon$,
the eigensolution for the normal displacement is given by

\begin{eqnarray}
	\xi_n(x) &=& \Xi_n(x) - \epsilon^2 \left[ \frac{x^2}{X^2} \ \Xi_n(x)
			+ \frac{x}{6} \left(1 - \frac{x^2}{X^2} \right) \Xi_n^\prime(x) \right]
			+ O(\epsilon^4) \; ,
\\ \nonumber \\
	\omega_n^2 &=& \kappa_n^2 c_0^2 \left[1 -
			\epsilon^2 \left( \frac{1}{3} + \frac{1}{2\kappa_n^2 X^2} \right) + O(\epsilon^4)\right] \; .
\label{eqn:eigfreq_norm}
\end{eqnarray}

By the same procedure the eigensolution for the binormal polarization can be obtained,

\begin{eqnarray}
	\zeta_n(x) &=& \xi_n(x) + O(\epsilon^4) \; ,
\\ \nonumber \\
	\omega_n^2 &=& \kappa_n^2 c_0^2 \left[1 -
			\epsilon^2 \left( \frac{1}{3} - \frac{1}{2\kappa_n^2 X^2} \right) + O(\epsilon^4)\right] \; .
\label{eqn:eigfreq_binorm}
\end{eqnarray}

\begin{figure*}%
        \epsscale{1.0}%
        \plotone{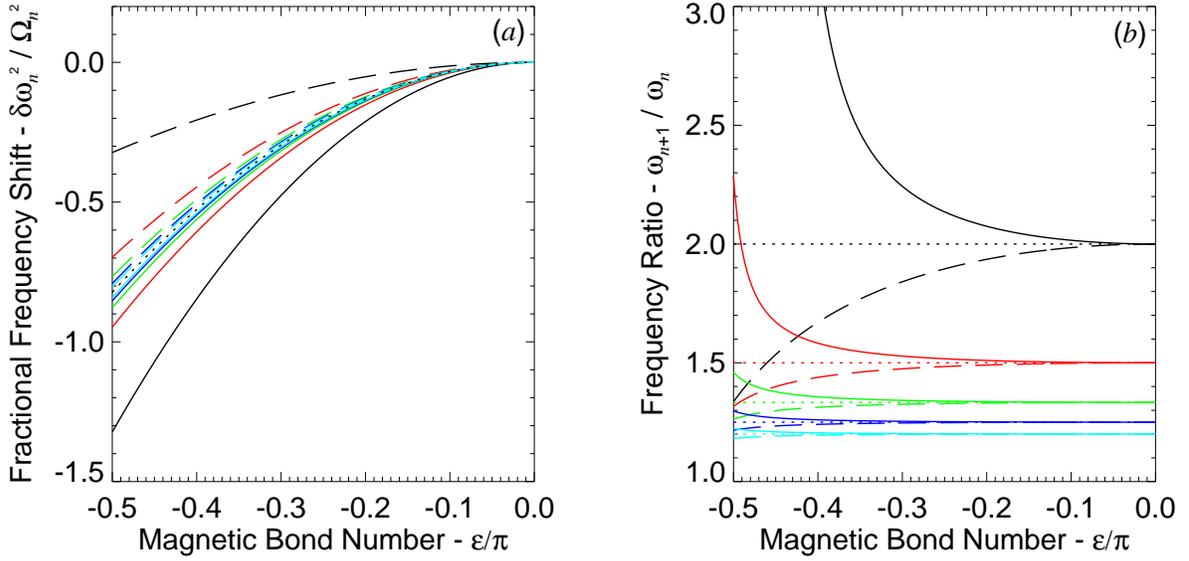}%
        \caption{\small The eigenfrequencies of coronal loop waves as a function
of the Magnetic Bond Number $\epsilon$. ($a$) Fractional perturbation in the square
eigenfrequency for both polarizations of motion. Transverse oscillations in the normal
direction are shown with the solid curves, while those in the binormal direction are
illustrated with dashed curves. The different colors correspond to different mode
orders: the fundamental ($n=1$) shown in black and the overtones in red ($n=2$),
green ($n=3$), blue ($n=4$), and aqua ($n=5$). The dotted curve corresponds to the
fractional frequency shift (independent of mode order) that arises solely from changes
in the length of the loop. ($b$) Frequency ratios of nearby mode orders, with the
same line styles and colors as above (i.e., black corresponds to $\omega_2/\omega_1$).
The dotted curves indicate the value appropriate for a straight loop in the absence
of gravity $(n+1)/n$. 
\label{fig:EigFreq}}%

\end{figure*}

\begin{figure*}%
        \epsscale{0.5}%
        \plotone{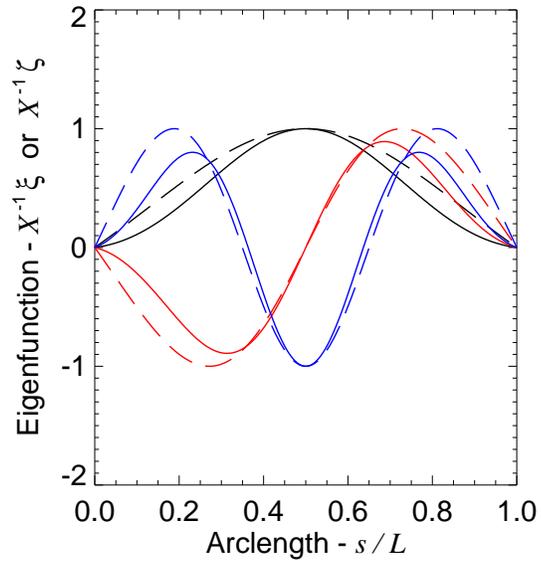}%
        \caption{\small Eigenfunctions for the fundamental (black) and first two
overtones (red and blue) as a function of arclength along the loop. The solid curves
illustrate the waveform for a Magnetic Bond Number of $\epsilon = -\pi / 4$, whereas the
dashed curves correspond to those for the unperturbed eigenfunction with $\epsilon = 0$.
Both displacement polarizations, $\xi$ and $\zeta$, possess the same eigenfunctions.
\label{fig:EigFunc}}%

\end{figure*}

\noindent Figure~\ref{fig:EigFreq} illustrates the resulting eigenfrequencies as
a function of $\epsilon$ for displacements in both the normal and binormal directions.
Figure~\ref{fig:EigFunc} shows sample eigenfunctions.

\section{Discussion}
\label{sec:Discussion}

In the previous sections we have modeled a coronal loop as an isolated, thin magnetic
flux tube. The effects of gravity and curvature have been self-consistently included,
leading to a static equilibrium determined by the balance of magnetic buoyancy and
tension. We have computed the eigensolutions for transverse oscillations of both
polarizations. In the following subsections we will discuss in detail the effects
that gravity and curvature have on the structure of the eigenfrequencies, assess
the differences between the two polarizations of wave motion, and explore the
implications of relaxing several of our operating assumptions.

\subsection{The Effects of Gravity and Curvature on the Eigenfrequencies}
\label{subsec:eigenfrequencies}

There are only two free parameters in our model, the footpoint separation $2X$
and the Magnetic Bond Number $\epsilon$.  The footpoint separation just provides
a physical scale for the problem and the structure of the solutions do not depend
upon it. The solutions do depend intrinsically on the Magnetic Bond Number, because
the equilibrium shape of the loop is sensitive to the strength of buoyancy and
magnetic tension. As the magnitude of the parameter $\epsilon$ becomes larger
(more negative), buoyancy forces increase in strength, which causes the loop to
rise and a corresponding increase in the curvature of the loop is needed to allow
magnetic tension to bring the loop back into equilibrium. This results in three
ways by which the frequencies are modified: the loop lengthens, the restoring
force of buoyancy increases, and the increased curvature modifies magnetic tension.
We can identify the first of these effects on the eigenfrequencies by expanding
the loop length $L$, equation~\eqnref{eqn:Length}, for small $\epsilon$,

\begin{equation}
	L = 2 X \left[ 1 + \frac{\epsilon^2}{6} + O(\epsilon^4) \right] \; .
\end{equation}

\noindent If the loop were straight and without gravitational effects, the
eigenfrequencies would be

\begin{equation}
	\omega_n^2 \sim \frac{n^2 \pi^2 c_0^2}{L^2} = \Omega_n^2
				\left[ 1 - \frac{\epsilon^2}{3} + O(\epsilon^4) \right] \; .
\end{equation}

\noindent As expected, an increase in the loop length (increasing $\epsilon^2$)
decreases the eigenfrequencies and the resulting fractional frequency shift is 
dispersionless and independent of mode order,

\begin{equation}
	\frac{\Delta \omega_n^2}{\Omega_n^2} = - \frac{\epsilon^2}{3} \; .
\end{equation}

\noindent This effect appears directly, in equations~\eqnref{eqn:eigfreq_norm}
and \eqnref{eqn:eigfreq_binorm} as the first term in the parentheses on the right-hand sides.

Buoyancy appears in identical form in the equations describing each polarization of
transverse oscillation; both equation~\eqnref{eqn:xieqn_raw} and \eqnref{eqn:zetaeqn_raw}
have the same first-derivative term. This common form, leads to identical perturbations
to the eigenfrequencies due to the inclusion of buoyancy. From equations~\eqnref{eqn:eigfreq_norm}
and \eqnref{eqn:eigfreq_binorm} we can directly see that buoyancy increases the
eigenfrequencies by the fractional amount,

\begin{equation}
	\frac{\Delta \omega_n^2}{\Omega_n^2} = \epsilon^2 \frac{2}{n^2 \pi^2} \; .
\end{equation}

\noindent This might be a somewhat surprising result. A simple dimensional analysis
of the equation of motion would indicate that buoyancy causes frequency shifts that
are first order in the Magnetic Bond Number $\epsilon$, and indeed this is what was
found in \cite{Jain:2012}. However, the buoyancy force in the wave equation is proportional
to $\epsilon$ times the geometrical factor $\unitv{z}\cdot\unitv{s}_0$. This geometrical
factor also depends on $\epsilon$ because the shape of the coronal loop is sensitive
to $\epsilon$. Low Magnetic Bond Numbers correspond to loops that are low-lying and nearly horizontal,
and because of this the geometrical factor is first order in the Magnetic Bond Number, and
the buoyancy force in total is second order.

The last effect, curvature of the loop, modifies the magnetic tension than can act as
a restoring force. It only appears in the equation for the normal polarization, and
its sign indicates that it counteracts the local curvature induced by the wave motions
(i.e., the second-derivative of the displacement). Therefore, we would expect
curvature of the background loop model to decrease the eigenfrequencies. This does
indeed occur, as we can see directly from the difference between the frequencies of
the normal and binormal polarizations. The curvature of the loop results in a
negative frequency shift that is twice as large in magnitude as the effect of buoyancy,

\begin{equation}
	\frac{\Delta \omega_n^2}{\Omega_n^2} = -\epsilon^2 \frac{4}{n^2 \pi^2} \; .
\end{equation}

The frequency shifts induced by both buoyancy and curvature enter at second order
in the Magnetic Bond Number. That both enter at the same order is a simple
consequence of the balance between buoyancy and tension in the equilibrium. That
they both enter at second order could easily be predicted. From dimensional arguments
alone, the shift in the square frequency due to curvature must scale with the radius
of curvature as $\Delta\omega^2 \sim c_0^2 R_0^{-2}$. Since the radius of curvature
is inversely proportional to $\epsilon$, the frequency shift must be second order
in $\epsilon$.

A consistent result for the curvature was found by \cite{vanDoorsselaere:2004} where
they solved for fast waves in a toroidal coordinate system in the absence of gravity.
The equilibrium field was assumed to be purely toroidal and a coronal loop modeled
by a partial section of the torus. With an extensive mathematical analysis they
concluded that the curvature of their torus induces a frequency shift that vanishes
at first order in the ratio of $a/R_0$, where $a$ is the minor radius of the torus
(and $R_0$ corresponds the the major radius). Therefore, the first non-zero perturbation
must occur at second or higher order. Here we have shown that for our thin tube model
the first non-zero perturbation to the frequency does indeed occur at second order
in the radius of curvature, as a simple dimensional analysis would indicate.

\subsection{The Two Polarizations of Oscillation}
\label{subsec:Polarizations}

Many of the observed transverse oscillations appear to be horizontal \cite[e.g.,][]{Aschwanden:2002},
and therefore in the binormal direction. Unfortunately, our assumption of an isolated
magnetic loop is particularly suspect for such oscillations. Most of what we call
coronal loops are probably just a bundle of field lines that are part of a much larger
magnetic structure like a magnetic arcade. We identify this particular bundle as a
``coronal loop" solely because it is selectively illuminated by a local heating process.
If the loop is part of a magnetic arcade, oscillations in the direction of the principle
normal (sometimes called ``vertical" oscillations) may be unaffected by the larger arcade
structure, because the undulations do not impinge on neighboring arcade field lines.
However, binormal (horizontal) oscillations must inherently impact nearby field lines
and the anisotropic nature of the arcade structure must modify wave propagation. In
particular, waves may no longer be confined to the loop. Spanwise propagation along
the arcade is possible, and since the other loops in the arcade may not be illuminated,
this propagation may go undetected. Such waves may still be trapped, but the problem
is inherently two-dimensional and our one-dimensional analysis incorrect.

Despite these difficulties, there are distinct differences between the two polarizations
of waves that may be observable. Unfortunately, the eigenfunctions are of little direct
use because the two polarizations have identical eigenfunctions (to second order in
the square of the Magnetic Bond Number). Therefore, observations of loop motion must be able
to directly account for projection and occultation effects if they are to determine
polarization. The frequencies of oscillation are not degenerate, however, and offer
promise to winnow between polarization modes.

Many observational and theoretical studies have examined the diagnostic value of the
ratio of eigenfrequencies \cite[for a review see][]{Andries:2009}, in particular the
ratio of the fundamental and first overtone. For a straight magnetic fibril in the
absence of gravity and stratification, the ratio of the first overtone to the fundamental
$\omega_2 / \omega_1$ is 2. Observational differences from 2 have been interpreted
as evidence for wave speed stratification \citep[e.g.,][]{McEwan:2008}, and potential
wave speed models have been rejected based on whether the model generates a ratio tha
is less than or greater than 2. As we shall soon see, curvature and buoyancy forces
can also cause the ratio to shift from the canonical value of 2, without any spatial
variation in the wave speed.

For our model, the frequency ratio for any given pair of neighboring modes is given by

\begin{equation}
	\frac{\omega_{n+1}}{\omega_n} = \frac{n+1}{n} \left[ 1 \pm \frac{\epsilon^2}{\pi^2}
				\frac{2n+1}{n^2(n+1)^2} + O(\epsilon^4)\right ] \; ,
\end{equation}

\noindent where the upper sign (+) corresponds to oscillations in the normal direction and
the lower sign ($-$) to those in the binormal direction. For the ratio of the first overtone
to the fundamental, this becomes 

\begin{equation}
	\frac{\omega_2}{\omega_1} = 2 \pm \frac{3}{2} \frac{\epsilon^2}{\pi^2} + O(\epsilon^4) \;.
\end{equation}

Since the effects of the variation in the loop length are dispersionless, they have dropped
out in these ratio expressions, leaving just the effects of buoyancy and curvature. Direct
buoyant augmentation of the restoring force causes the ratio to decrease, while curvature
of the loop causes the ratio to increase. Since the normal polarization senses the effects
of loop curvature, whereas the binormal polarization does not, the ratio of the first overtone
to the fundamental is greater than 2 for normal oscillations and less than 2 for binormal
displacements. The diagnostic value of the frequency ratio is clearly predicated on the
knowledge of which polarization one is observing.

\subsection{The Effects of an External Magnetic Field}
\label{subsec:Bext}

While we have assumed that the coronal loop is embedded in a field-free atmosphere,
we can speculate what the effects of an external magnetic field might be. Let's first
consider the equilibrium. The inclusion of an external magnetic field only provides
minor modifications to the steady state form of the equation of motion~\eqnref{eqn:Spruit}.
The tangential component, equation~\eqnref{eqn:equil_s}, remains unchanged as it
depends only on internal fluid properties. The same can be argued for the magnetic
tension as long as the thin-tube approximation holds. The buoyancy force is modified,
however, because the density difference between the internal and external fluid is a
function of the external magnetic field through equilibration of the total pressure
across the tube. However, if the external field is force-free, the density scale height
of the external fluid is unaffected and density changes only manifest as a different
value of the Magnetic Bond Number $\epsilon$---see equation~\eqnref{eqn:epsilon}.
Therefore, the equation that describes the equilibrium shape of the loop,
equation~\eqnref{eqn:EqODE}, remains valid in the presence of external field, and of
course the solutions of that equation also remain valid.

The existence of an external magnetic field is not as benign for the wave equations
satisfied by the kink waves. The presence of the surrounding magnetic field can be
felt by the waves through backreaction forces caused by the tube's motion. The well-known
effect of enhanced inertia, which accounts for the fact that a moving tube must move
external fluid out of its way, has already been included in our equation of
motion~\eqnref{eqn:Spruit}. However, when the external fluid is magnetized, there
is an additional effect because bending of the tube requires bending of the external
field as well. Therefore, the external field $B_\e$ provides additional stiffness that
increases the kink wave speed,

\begin{equation}
	c_K^2 = \frac{B^2 + B_\e^2}{4\pi (\rho + \rho_\e)} \; .
\end{equation}

\noindent This well-known expression for the kink wave speed  \citep[e.g.,][]{Edwin:1983}
displays both backreaction forces explicitly. In the denominator the sum of the internal
and external densities is the effect of the enhanced inertia, while in the numerator
the appearance of $B_\e^2$ is the result of the increases stiffness. These forces are
derived and identified explicitly in Appendix~\ref{app:Backreaction} for a thin tube
in the absence of gravity.

When including the effects of an external magnetic field, one is tempted to simply
replace the kink speed $c_0$ appearing in the wave equations~\eqnref{eqn:xieqn_raw}
and \eqnref{eqn:zetaeqn_raw} with the proper wave speed $c_K$,

\begin{eqnarray}
\label{eqn:xieqn_Be_raw}
	\dern[\xi]{s}{2} - \frac{g_\parallel D_0}{c_K^2} \der[\xi]{s}
		+ \left(\frac{\omega^2}{c_K^2} + \frac{1}{R_0^2}\right) \xi &=& 0 \; ,
\\ \nonumber \\
	\dern[\zeta]{s}{2} - \frac{g_\parallel D_0}{c_K^2} \der[\zeta]{s}
		+ \frac{\omega^2}{c_K^2} \zeta &=& 0 \; .
\label{eqn:zetaeqn_Be_raw}
\end{eqnarray}
	
\noindent From the motivating calculation in Appendix~\ref{app:Backreaction} it is
evident that such a substitution is justified in the absence of curvature of the loop.
When curvature is present, the validity of the substitution seems physically appropriate,
but remains mathematically unproven. While these two wave equations possess the same
form as those solved previously, equations~\eqnref{eqn:xieqn_raw} and \eqnref{eqn:zetaeqn_raw},
there is one subtle difference. The ratio appearing in front of the first derivative
in each equation  is no longer related to the parameter $\epsilon$ in the same fashion.
Previously, the denominator depended on $c_0$ instead of $c_K$. Since, the magnetic
tension in the equilibrium still depends on $c_0$ (and not $c_K$), after inserting
$g_\parallel$ appropriate for the equilibirum, we find

\begin{eqnarray}
\label{eqn:xieqn_Be}
	\dern[\xi]{s}{2} + \alpha \frac{\epsilon}{X} \sin(\epsilon x/X) \, \der[\xi]{s}
		+ \left(\frac{\omega^2}{c_K^2} + \frac{1}{R_0^2}\right) \xi &=& 0 \; ,
\\ \nonumber \\
	\dern[\zeta]{s}{2} + \alpha \frac{\epsilon}{X} \sin(\epsilon x/X) \, \der[\zeta]{s}
		+ \frac{\omega^2}{c_K^2} \zeta &=& 0 \; ,
\label{eqn:zetaeqn_Be}
\end{eqnarray}

\noindent with $\alpha$ being a factor that depends on the plasma parameter within the
tube, $\beta$, and outside the tube, $\beta_\e$,

\begin{equation}
	\alpha \equiv \frac{B^2}{B^2 + B_\e^2} = \frac{1+ \beta_\e}{2 + \beta + \beta_\e} \; .
\end{equation}

\noindent Since the corona is magnetically dominated  (i.e., $\beta<<1$ and $\beta_e<<1$),
$\alpha$ is generally quite close to a value of one-half.  The form of our previous wave
equations can be recovered only for $\alpha = 1$.  While this seems like a minor difference,
the change of variable that we performed to remove the first derivatives in equations~\eqnref{eqn:xieqn}
and \eqnref{eqn:zetaeqn} is predicated on $\alpha = 1$. So, while the basic form of
the wave equations remains unchanged when we include an external magnetic field, the
detailed perturbation analysis performed in \S\ref{sec:KinkWaves} is no longer valid.
It would be relatively trivial to repeat the perturbation analysis for this new case,
but we choose to leave such an effort to a future study.

\section{Conclusions}
\label{sec:Conclusions}

We have developed a model of a coronal loop that self-consistently includes the
effects of gravity and curvature. This has been accomplished for a loop that
is treated as an isolated magnetic fibril embedded in an isothermal field-free
corona. For such a model, the shape adopted by the loop when in a state of static
equilibrium is completely determined by a balance between the forces of buoyancy
and magnetic tension. The height of the loop depends on the Magnetic Bond
Number $\epsilon$ which is usually small for coronal loops. For these small values,
the loops are low-lying with weak curvature. In the other extreme, no stable
equilibrium exists for $|\epsilon|>\pi/2$. This stability criterion is actually
a restriction on the footpoint separation compared to the coronal density scale
height. Loops with a footpoint separation greater than $2 \pi H$ are unstable.

Since the static loop is confined within a vertical plane, and only possesses
curvature within this plane, the two polarizations of transverse oscillations
(those in the direction of the principle normal and those in the direction of
the binormal) decouple. The wave equations that describe these oscillations
both have restoring forces from buoyancy and from the magnetic tension induced
by undulations. The equation that describes the normal polarization also possesses
a tension term that arises from the curvature of the loop itself.  The effects
of the restoring force of buoyancy and of background curvature both enter the
eigenfrequencies at second order in the Magnetic Bond Number. Buoyancy
increases the frequency and curvature reduces it. The ratio of the frequencies
of two eigenmodes of neighboring order $\omega_{n+1}/\omega_n$ differ from the
canonical value of $(n+1)/n$ that would be achieved for a straight loop in the
absence of gravity with constant kink speed. Oscillations confined to the plane
of the loop, or those polarized in the normal direction, have a frequency ratio
that exceeds this canonical value, $\omega_{n+1}/\omega_n > (n+1)/n$, whereas
horizontal oscillations, or the binormal polarization, have a ratio that is less
than the canonical value, $\omega_{n+1}/\omega_n < (n+1)/n$. These conditions
on the frequency ratios are achieved for a loop with a constant kink wave speed,
independent of the pressure scale height.


\acknowledgements

This work is supported by NASA, MSRC (University of Sheffield) and STFC (UK). BWH
acknowledges NASA grants NNX08AJ08G, NNX08AQ28G, and NNX09AB04G.

\appendix\section{Backreaction Forces on a Magnetized Tube in a Magnetized Environment}
\label{app:Backreaction}

In order to see how the backreaction forces might appear for a magnetic fibril
moving through a {\sl magnetized} external fluid, we perform a simple wave emission
calculation for a field configuration amenable to analytical treatment and direct
identification of the relevant forces. Following \cite{Edwin:1983}, consider a magnetized
cylinder embedded in a magnetized atmosphere in the absence of gravity. The magnetic
field within the cylinder is constant and points in the $z$-direction,
$\bvec{B} = B \unitv{z}$.  The magnetic field in the region exterior to the cylinder
is also constant and points in the same direction $\bvec{B}_\e = B_\e \unitv{z}$.
While the two fields are aligned, they may potentially differ in strength. The
mass densities inside and outside the cylinder are $\rho$ and $\rho_\e$, respectively. 

Magnetosonic waves outside the cylinder have a simple analytic form when expressed
in polar coordinates with the axis of the coordinate system coaligned with the axis
of the cylinder. A general form for the radial fluid displacement $\xi_r$ and the
total pressure perturbation $\delta \Pi$ are as follows \citep{Edwin:1983}:

\begin{eqnarray}
	\xi_r &=& \kappa^{-1} e^{im\phi} \ Q_m^\prime\left(\lambda r\right) \ e^{i \kappa z} \ e^{-i \omega t} \; ,
\\ \nonumber \\
	\delta \Pi &=& \rho_e \, \left(\lambda \kappa\right)^{-1} \left(\omega^2 - \kappa^2 V_e^2\right) \
		e^{im\phi} \ Q_m\left(\lambda r\right)  \ e^{i \kappa z} \ e^{-i \omega t} \; ,
\end{eqnarray}

\noindent with the dispersion relationship

\begin{equation}
	\lambda^2 = \frac{\left(\omega^2 - \kappa^2 c_e^2\right) \left(\omega^2 - \kappa^2 V_e^2\right)}
			{ \left(c_e^2 + V_e^2\right) \left(\omega^2 - \kappa^2 U_e^2\right) } \; .
\end{equation}

\noindent In these equations, $c_e$, $V_e$ and $U_e$ correspond to the external
values of the sound, Alfv\'en, and slow wave speeds, respectively. The axial and
radial wavenumbers are denoted $\kappa$ and $\lambda$, and the temporal frequency is
$\omega$. The function $Q_m$ is a $m$th-order Bessel function or Hankel function.
Primes denote differentiation with respect to the argument of the Bessel
function.

If an axially propagating wave causes the tube to oscillate back and forth in the
$\unitv{x}$ direction with an amplitude $\eta$,

\begin{equation}
	\bvec{\xi}_{\rm tube} = \eta  \ e^{i \kappa z} \ e^{-i \omega t} \unitv{x} \; ,
\end{equation}

\noindent the radial displacement at the lateral surface of the tube ($r = a$) is given by

\begin{equation} \label{eqn:xi_r}
	\xi_{r,{\rm tube}} = \eta \cos\phi  \ e^{i \kappa z} \ e^{-i \omega t}\; .
\end{equation}

\noindent The radial displacement of the external fluid must match the normal
motion of the tube's outer surface. Therefore, the wavefield in
the external medium must have the same $\cos\phi$ azimuthal dependence as appears
in equation~\eqnref{eqn:xi_r}  and will be composed of only the dipole components
($m = \pm 1$). Further if $\lambda^2 > 0$, causality requires that the wavefield
only possesses outward propagating waves, leading to the following form:

\begin{eqnarray}
\label{eqn:xi_e}
	\xi_r &=& A \kappa^{-1} \cos\phi \ H_1^\prime\left(\lambda r\right) \ e^{i \kappa z} \ e^{-i \omega t} \; ,
\\ \nonumber \\
	\delta \Pi &=& A \kappa^{-1} \rho_e \ \lambda^{-1} \left(\omega^2 - \kappa^2 V_e^2\right) \
		\cos\phi \ H_1\left(\lambda r\right)  \ e^{i \kappa z} \ e^{-i \omega t} \; ,
\label{eqn:Pi_e}
\end{eqnarray}

\noindent with the function $H_1$ corresponding to an outward propagating Hankel
function of the first kind, $H_1(\lambda r) = H_1^{(1)}(\lambda r)$, and the wave
amplitude $A$ fixed by the displacement matching condition at the tube's surface,

\begin{equation} \label{eqn:Acoef}
	A = \frac{\kappa\eta}{H_1^\prime\left(\lambda a\right)} \; .
\end{equation}

The backreaction forces acting on the tube due to the tube's motion through the
external fluid can be derived by integrating the forces acting on the tube
over the tube's cross-section. In the absence of gravity the only forces acting
on the tube are magnetic tension and total pressure. If the tube is thin,
we may ignore internal variations of the magnetic field strength $B$ and the cross-sectional
mean of the tension force acting on the tube is given by

\begin{equation}
	\bvec{F}_{\rm tension} = \frac{B^2}{4\pi} \dern[\bvec{\xi}_{\rm tube}]{z}{2}
			= -\rho V^2 \kappa^2 \eta \ e^{i \kappa z} \ e^{-i \omega t} \, \unitv{x} \; ,
\end{equation}

\noindent where $V$ is the Alfv\'en speed within the tube.

The cross-sectional mean of the total pressure force can be expressed as the integral
of the total pressure perturbation around the circumference of the tube by using
Stokes theorem to convert the area integral into a contour integral,

\begin{equation}
	\bvec{F}_{\rm pressure} = -\frac{1}{\pi a^2} \int_0^{2\pi} \delta\Pi(a,\phi,z) \, \unitv{r} \ a \, d\phi \; .
\end{equation}

\noindent The $x$-component of the total pressure force is the only nonvanishing
component, and the mean force in this direction is therefore given by the sum of
the pressure and tension forces,

\begin{equation}
	F_x = -\frac{1}{\pi a} \int_0^{2\pi} \cos\phi \, \delta\Pi(a,\phi,z) \ d\phi 
		-\rho V^2 \kappa^2 \eta \ e^{i \kappa z} \ e^{-i \omega t} \; .
\end{equation}

The integral appearing in this equation is trivial because the total pressure
field on the surface of the tube only has dipole dependence. Using
equations~\eqnref{eqn:Pi_e} and \eqnref{eqn:Acoef} this reduces to

\begin{equation}
	F_x = -\left[\rho \kappa^2 V^2 \eta + \rho_e \left(\omega^2 - \kappa^2 V_e^2\right) \,
		(\lambda a)^{-1} \frac{H_1(\lambda a)}{H_1^\prime(\lambda a)} \ \eta \right]
		\ e^{i \kappa z} \ e^{-i \omega t} \; .
\end{equation}

\noindent Since the tube is thin we can perform small argument expansions on
the Hankel functions to obtain a simple form,

\begin{equation}
	F_x = \left[ -\kappa^2 \frac{B^2}{4\pi} \eta
		+ \left(\rho_e \omega^2 - \kappa^2 \frac{B_e^2}{4\pi}\right) \eta \right]
		\ e^{i \kappa z} \ e^{-i \omega t} \; .
\end{equation}

The equation of motion can therefore be written

\begin{equation}
	-\rho \omega^2 \bvec{\xi}_{\rm tube} = \frac{B^2+B_e^2}{4\pi} \dern[]{z}{2}\bvec{\xi}_{\rm tube}
		+ \rho_e \omega^2 \bvec{\xi}_{\rm tube} \; .
\end{equation}

We can immediately see that there are two backreaction forces. Both arise from
the total pressure in the external fluid. The term proportional to $\rho_e\omega^2 \bvec{\xi}_{\rm tube}$
is the familiar effect of enhanced inertia and arises from the $\omega^2$ factor
appearing in equation~\eqnref{eqn:Pi_e}. In the absence of external field, this
would be the only term.  However, when external field is present, the total
pressure in the external fluid is reduced by the fractional amount
$\left(\omega^2-\kappa^2 V_e^2\right)/\omega^2$ compared to pressure in the unmagnetized
fluid. The term proportional to the square of the Alfv\'en speed leads to an
additional backreaction term $-\kappa^2 (B_\e^2 / 4 \pi) \bvec{\xi}_{\rm tube}$
that represents the force needed to bend the external field lines when the tube
itself bends.  The external field provides additional stiffness that increases the
wave speed.




\end{document}